# Experimental Study of Electron and Phonon Dynamics in Nanoscale Materials by Ultrafast Laser Time-Domain Spectroscopy

by
Xiaohan Shen
A Dissertation Submitted to the Graduate
Faculty of Rensselaer Polytechnic Institute
in Partial Fulfillment of the
Requirements for the degree of
DOCTOR OF PHILOSOPHY
Major Subject: Physics

Approved by the
Examining Committee:

_________________________________________

Prof. Masashi Yamaguchi, Thesis Advisor

_________________________________________

Prof. Toh-Ming Lu, Member

_________________________________________

Prof. Shawn-Yu Lin, Member

_________________________________________

Prof. Kim M. Lewis, Member

_________________________________________

Prof. Ganpati Ramanath, Member

# CONTENTS

## LIST OF TABLES

# LIST OF FIGURES

# ACKNOWLEDGEMENT

First of all, I would like to express my gratitude and appreciation to my advisor, Professor Masashi Yamaguchi for his support and guidance in my graduate career. He provided me the opportunity to work with the femtosecond high power laser system and guided me step by step into the optical experiment. He also taught me how to presentation skills, such as slides preparation and paper writing. His insistence on high standards and strong attention to details inspired me to improve constantly the experiments as well as my research.

I would like to thank all the members of my dissertation committee, Professor Toh-Ming Lu, Professor Shawn-Yu Lin, Professor Kim M. Lewis, and Professor Ganpati Ramanath for reviewing and evaluating my work. Their insightful comments helped me to refine my physical understanding and further my studies.

I am thankful to my previous colleague Dr. Zhengping Fu in our group. Discussion with him was enjoyable and helpful for the advances in both experiments and analysis. It has been a great pleasure to work with him in the past few years. I also want to take this opportunity to thank the past members in our group, Mr. Matthew Grabala and Mr. Zachary Boak.

I sincerely acknowledge Dr. Zonghuan Lu, Dr. Yukta P. Timalsina, Dr. Andrey Muraviev, Dr. Kamaraju Natarajan, Dr. Andrej Halabica, Dr. Thomas Cardinal, Mr. Matthew Kwan, and Dr. Peter O’Brien for their collaboration and help.

Last but not the least, I am greatly indebted to my parents for their consistent support and encouragement. This dissertation is dedicated to you.

## LIST OF JOURNAL PUBLICATIONS

1. **Xiaohan Shen**, Zonghuan Lu, Yukta P. Timalsina, Toh-Ming Lu, and Masashi Yamaguchi, “Direct measurement of phonon mean-free-path and control of acoustic excitation using frequency and spectral width tunable narrowband acoustic phonon spectroscopy”, (2017). To be submitted

2. Thomas Cardinal, Matthew Kwan, **Xiaohan Shen**, Masashi Yamaguchi, Theodorian Borca-Tasciuc, and Ganpati Ramanath, “Increasing thermoelectric device efficiency by nanomolecular functionalization of metal-thermoelectric interfaces”, J. Appl. Phys. (2017). Submitted

3. M. Shur, S. Rudin, G. Rupper, M. Yamaguchi, **X. Shen**, A. Muraviev, “Subpicosecond nonlinear plasmonic response probed by femtosecond optical pulses”, Int. J. High Speed Electron. Syst. Vol. **25**, Nos. 1&2 (2016) 1640003.

4. A. Muraviev, A. Gutin, G. Rupper, S. Rudin, **X. Shen**, M. Yamaguchi, G. Aizin, M. Shur, “ New optical gating technique for detection of electric field waveforms with subpicosecond resolution”, Opt. Express **24**, 12730-12739 (2016).

5. **Xiaohan Shen**, Yukta P. Timalsina, Toh-Ming Lu, Masashi Yamaguchi, “Experimental study of electron-phonon coupling and electron internal thermalization in epitaxially grown ultrathin copper films”, Phys. Rev. B **91**, 045129 (2015).

6. Yukta P. Timalsina, **Xiaohan Shen**, Grant Boruchowitz, Zhengping Fu, Guoguang Qian, Masashi Yamaguchi, Gwo-Ching Wang, Kim M. Lewis, Toh-Ming Lu, “Evidence of enhanced electron-phonon coupling in ultrathin epitaxial copper films”, Appl. Phys. Lett. **103**, 191602 (2013).

## PRESENTATIONS AND CONFERENCE PUBLICATIONS

1. **Xiaohan Shen**, Yukta P. Timalsina, Toh-Ming Lu, Masashi Yamaguchi, Fall Meeting of Hudson Mohawk chapter of the AVS, Topic: “Experimental study of spatial confinement effect on acoustic phonon transport in $SiO_2$ nanorods using ultrafast pump-probe spectroscopy”, Union College, Schenectady, NY, September, 2015.

2. A. Gutin, A. V. Muraviev, N. Kamaraju, **X. Shen**, M. Yamaguchi, M.S. Shur, D. But, N. Dyakonova, and W. Knap, $39^{th}$ Int. Conf. on Infrared, Millimeter, and THz waves, Topic: “Application of plasma-wave detectors for ultra-short pulse terahertz radiation”, The University of Arizona, Tucson, AZ, September, 2014.

3. **Xiaohan Shen**, Zhengping Fu, Masashi Yamaguchi, Yukta P. Timalsina, Toh-Ming Lu, MRS Fall Meeting & Exhibit, Topic: “Ultrafast optical study of electron-phonon coupling in epitaxially grown copper films”, Boston, MA, December, 2013.

4. **Xiaohan Shen**, Yukta Timalsina, Toh-Ming Lu, Masashi Yamaguchi, School of Science Graduate Symposium, Topic: “Femtosecond pump-probe study of size effect on electron-phonon coupling in copper films”, Rensselaer Polytechnic Institute, Troy, NY, April, 2013.

5. **Xiaohan Shen**, Andrej Halabica, Pei-I Wang, Toh-Ming Lu, Masashi Yamaguchi, Spring Meeting of Hudson Mohawk chapter of the AVS, “Femtosecond pump-probe study of electron-phonon coupling in copper thin films”, Rensselaer Polytechnic Institute, Troy, NY, April, 2012.

# ABSTRACT

With the rapid advances in the development of nanotechnology, nowadays, the sizes of elementary unit, i.e. transistor, of micro- and nanoelectronic devices are well deep into nanoscale. For the pursuit of cheaper and faster nanoscale electronic devices, the size of transistors keeps scaling down. As the miniaturization of the nanoelectronic devices, the electrical resistivity increases dramatically, resulting rapid growth in the heat generation. The heat generation and limited thermal dissipation in nanoscale materials have become a critical problem in the development of the next generation nanoelectronic devices. Copper (Cu) is widely used conducting material in nanoelectronic devices, and the electron-phonon scattering is the dominant contributor to the resistivity in Cu nanowires at room temperature. Meanwhile, phonons are the main carriers of heat in insulators, intrinsic and lightly doped semiconductors. The thermal transport is an ensemble of phonon transport, which strongly depends on the phonon frequency. In addition, the phonon transport in nanoscale materials can behave fundamentally different than in bulk materials, because of the spatial confinement. However, the size effect on electron-phonon scattering and frequency dependent phonon transport in nanoscale materials remain largely unexplored, due to the lack of suitable experimental techniques.

This thesis is mainly focusing on the study of carrier dynamics and acoustic phonon transport in nanoscale materials. The weak photothermal interaction in Cu makes thermoreflectance measurement difficult, we rather measured the reflectivity change of Cu induced by absorption variation. We have developed a method to separately measure the processes of electron-electron scattering and electron-phonon scattering in epitaxial Cu films by monitoring the transient reflectivity signal using the resonant probe with particular wavelengths. The enhancement on electron-phonon scattering in epitaxial Cu films with thickness less than 100 nm was observed. The longitudinal acoustic phonon transport in silicon (Si) nanorod with confined diameter and length was investigated. The guided phonon modes in Si nanorod with different frequencies and wave vectors were observed. The mean-free-path of the guided phonons in Si nanorod was found to be larger than the effective phonon mean-free-path in Si film, because of the limited phonon scattering channels in Si nanorod. The phonon density of states and dispersion relation strongly depend on the size and boundary conditions of nanorod. Our work demonstrates the possibility of modifying the phonon transport properties in nanoscale materials by designing the size and boundary conditions, hence the control of thermal conductivity. In addition, the periodicity effect of nanostructures on acoustic phonon transport was investigated in silicon dioxide ($SiO_2$) nanorod arrays. The lattice modes and mechanical eigenmodes were observed, and the pitch effect on lattice modes was discussed. A narrowband acoustic phonon spectroscopic technique with tunable frequency and spectral width throughout GHz frequency range has been developed to investigate the frequency-dependent acoustic phonon transport in nanoscale materials. The quadratic frequency dependence of acoustic attenuation of $SiO_2$ and indium tin oxide (ITO) thin films was observed, and the acoustic attenuation of ITO was found to be larger than $SiO_2$. Moreover, the acoustic control on mechanical resonance of nanoscale materials using the narrowband acoustic phonon source was demonstrated in tungsten thin film.